\documentclass[prb,twocolumn,showpacs]{revtex4}

\usepackage{graphicx}
\usepackage{amsmath}
\usepackage{amssymb}
\usepackage{epsf,latexsym}
\usepackage{times}

\newcommand{\ket}[1]{\left| #1 \right\rangle}
\newcommand{\bra}[1]{\left\langle #1 \right|}
\newcommand{\be}{\begin{equation}}
\newcommand{\ee}{\end{equation}}
\newcommand{\ba}{\begin{eqnarray}}
\newcommand{\ea}{\end{eqnarray}}
\newcommand{\bc}{\begin{center}}
\newcommand{\ec}{\end{center}}

\begin{document}


\title{Creating, distributing and freezing entanglement with spin chains}

\author{Irene D'Amico$^{1}$}
\email{ida500@york.ac.uk}
\author{Brendon W. Lovett$^{2}$}
\email{brendon.lovett@materials.oxford.ac.uk}
\author{Timothy P. Spiller$^{3}$}
\email{tim.spiller@hp.com}

\affiliation{
$^1$ Department of Physics, University of York, York YO10 5DD, United Kingdom\\
$^2$ Department of Materials, University of Oxford, OX1 3PH, United Kingdom\\
$^3$ Hewlett-Packard Laboratories, Filton Road,
Stoke Gifford, Bristol BS34 8QZ, United Kingdom}

\date{\today }


\date{\today }

\begin{abstract}
We show how branching spin chains can be used to both generate and
distribute entanglement from their natural dynamics. Such
entanglement provides a useful resource, for example for
teleportation or distributed quantum processing. Once distributed,
this resource can be isolated through mapping or swapping the entanglement into
specific qubits at the ends of branches. Alternatively, as we demonstrate for
distributed bipartite entanglement, applying simple single-qubit
operations to the end spin of one or more branches can dynamically freeze the
entanglement at the branch ends.

\end{abstract}




\maketitle

\section{Introduction}
Over the past few years there has been significant interest in the
propagation of quantum information through spin chains. At a
fundamental level, it is interesting to determine just what the
natural dynamics of these systems permits. However, from a more
practical quantum technology perspective, the controlled propagation
of quantum information has potential use. For macroscopic distances,
the current consensus is that photons, or other quantum states of
light, form the best medium for quantum communication, as these can
propagate through optical fibres or free space with high fidelity
\cite{qkd}. However, for much shorter, microscopic, distances it is
possible that other media, such as spin chains, could make a very
useful contribution. For example, in future solid state quantum
information devices there could be a need to provide quantum
communication links over microscopic distances, between separate
quantum processors or registers, or between processors and memory,
analogous to the conventional communication that goes on within the
computer chips of today.

In this paper we use the terminology `spin chain' to cover any
physical system whose dynamics can be predicted using a Hamiltonian
isomorphic to that of a coupled spin chain, which we shall define
precisely later. In practice, this could be strings of actual spins
(produced chemically, or fabricated) connected through interactions,
or it could be a string of quantum dots or molecules (like
fullerenes), containing exciton or spin qubits. Strings of trapped
atoms form another possibility.

In order to quantify how well quantum processors or registers might
communicate, there has been an emphasis on how well a quantum state
can propagate through a spin chain or network. Originally it was
shown that a single spin qubit could transfer with decent fidelity
along a constant nearest-neighbour exchange-coupled chain
\cite{bose03}. The fidelity can approach unity if the qubit is
encoded into a packet of spins \cite{osb04}. Perfect state transfer
can also be achieved in more complicated systems, with different
geometry or chosen unequal couplings
\cite{chr04,chr05,dechi05,yun05,kar05}. Parallel spins chains also enable
perfect transfer \cite{bur05a,bur05b}, as does a chain used as a
wire, with controlled couplings at the ends \cite{woj05}. Related
studies have also been made on chains of quantum dots
\cite{damico05}, chains of quantum oscillators \cite{ple05,har05}
and spin chains connected through long range magnetic dipole
interactions \cite{avel06}. State transfer and operations through
spin chains using adiabatic dark passage has also been proposed
\cite{gre05,ohs07}.

In this work,  we consider a different approach to the problem of
short distance quantum communication. Rather than focus on spin
chains providing high fidelity transfer of quantum information, our
approach is to construct spin chain systems whose natural dynamics
creates high fidelity spatially separated entanglement from a simple
initial state. We will show that this can be achieved in 
branched spin chain structures with certain coupling strengths. 
The consequences of branching were first studied in the context of 
divided {\it bosonic} chains, composed of coupled harmonic 
oscillators~\cite{plenio04, perales05} - and have 
since been studied in systems of propagating electrons~\cite{yang06, yang07}. In both of 
these cases, the dynamics of Gaussian wave packet type excitations have 
been investigated. By contrast, our paper deals with the propagation of 
a single spin-up excitation localized on a single site, and how this 
propagates through a network of spin-down states.

We shall further introduce a technique for {\it freezing} dynamically 
generated entanglement.
Clearly such entanglement is a useful and flexible resource
\cite{gre05,plenio04, perales05,yang06, yang07,dev05,pat05,spi06} for
quantum information. For example, it could be used for
quantum teleportation \cite{tele93}, which enables the transfer of a
quantum state, or to connect separated quantum registers or
processors.  One advantage of this approach is that decoherence or
imperfections, which can lead to imperfect entanglement production,
could be countered by purification techniques \cite{pur96} prior to
use of the entangled resource.

\section{Spin chain formalism}
To introduce our formalism, we first consider a one-dimensional
chain of $N$ spins, each coupled to their nearest neighbours. The
Hamiltonian for the system is
\ba
H &=& -\sum_{i=1}^N \frac{E_i}{2} \sigma_{z}^{i} \nonumber \\
&+&
 \sum_{i=1}^{N-1} \frac{J_{i,i+1}}{2}\left(\sigma_{+}^{i}\sigma_{-}^{i+1} + \sigma_{-}^{i}\sigma_{+}^{i+1}\right)
 \;,
 \label{Hspin}
\ea where $\sigma_{z}^{i}$ is the $z$ Pauli spin matrix for the spin
at site $i$ and similarly $\sigma_{\pm}^i=\sigma_{x}^{i} \pm i
\sigma_{y}^{i}$. For actual spins, $E_{i}/2$ is the local magnetic
field (in the $z$-direction) at site $i$ and $J_{i,i+1}/2$ is the
local $XY$ coupling strength between neighbouring sites $i$ and
$i+1$. For a coupled chain of quantum dots where each qubit is
represented by the presence or absence of a ground state exciton,
$E_i$ is the exciton energy and $J_{i,i+1}$ is the F\"orster
coupling between dots $i$ and $i+1$ \cite{damico05}. The spin chain
formalism applies to both such physical systems, and to any others
which can be described by an isomorphic Hamiltonian.
The computational basis notation for the spin
states at each site is $|0\rangle_i \equiv |\uparrow_z\rangle_i$ and
$|1\rangle_i \equiv |\downarrow_z\rangle_i$.

The total $z$-component of spin (magnetization), or total exciton
number, is a constant of motion as it commutes with $H$. It is
therefore instructive
to consider a state of the system consisting of the ground
state with the addition of a single flipped spin. This state is
straightforward to prepare, assuming local control over a spin at,
say, the end of a chain, and the flipped spin can be regarding as a
(conserved) travelling qubit as it moves around under the action of the chain
dynamics \cite{bose03}. An efficient way of representing
such states in an $N$ spin network is the site basis defined as
$\ket{\mathsf  k} = \ket{0_1, 0_2, ..., 0_{k-1}, 1_k, 0_{k+1}, ..., 0_N}$.
A system prepared in this subspace remains in it. Now the detailed dynamics
depend on the local magnetic fields or exciton energies, but if these
are independent of location $i$ then the dynamics favour quantum state
transfer processes, as already mentioned. We also adopt this limit
for our work here.

\section{Bi-partite entanglement creation and distribution}
\label{bipartite}
\begin{figure}
\bc
\includegraphics[scale=0.25]{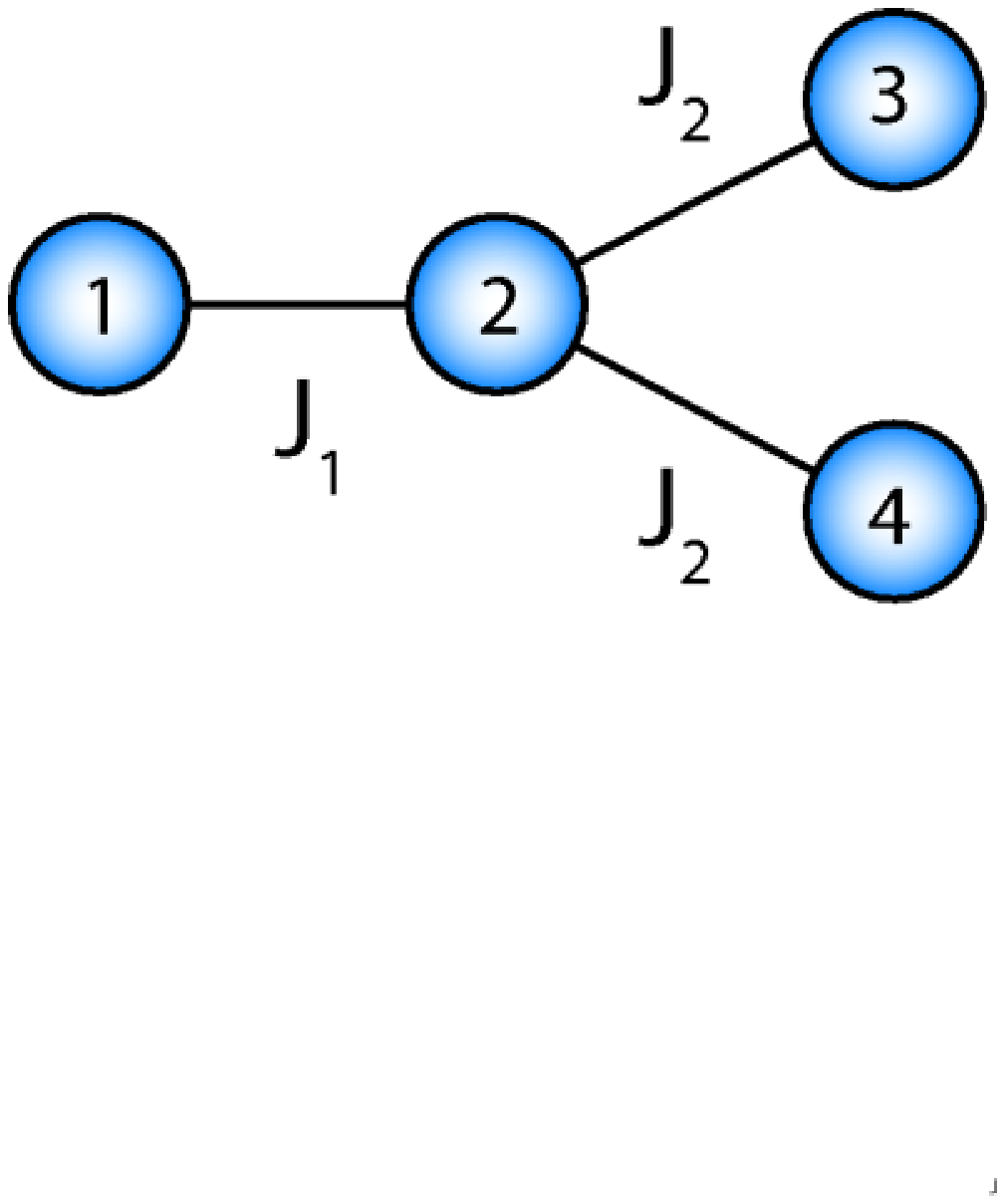}
\vspace{-2cm}
\caption{A schematic four-site Y spin chain system, with nearest
neighbour couplings as shown.} \label{fig1}
\ec\end{figure}

Rather than using spin chains to transfer quantum states, our focus
here is to use spin chain systems to prepare spatially separated
entanglement, which can then be used as a resource for
teleportation, or providing quantum connections between separated
quantum processors or registers. The simplest system we consider is
a Y structure, chosen to prepare bi-partite entanglement from a
simple initial state. The smallest definable Y structure contains
four sites, which we label as shown in figure~\ref{fig1}.

As seen in the figure, there is a deliberate asymmetry in the
coupling of the outside sites to the hub of the Y. There is one
``input'' site, coupled to the hub with strength $J_1$, and two ``output''
sites, coupled to the hub with equal strength $J_2$.
The Hamiltonian for this system is written:
\be
H_{Y4} = J_1\ket{\mathsf 1}\bra{\mathsf  2}+J_2(\ket{\mathsf  2}\bra{\mathsf  3}+\ket{\mathsf  2}\bra{\mathsf  4})+H.c. .
\ee
We proceed by noting that $\ket{-}\equiv2^{-\frac{1}{2}}(\ket{\mathsf  3}-\ket{\mathsf  4)}$ is an eigenstate of the system. If we initialize our system in the state $\ket{\mathsf  1}$ then $\ket{-}$ is decoupled and plays no part in our system dynamics. We define the orthogonal state $\ket{+}\equiv 2^{-\frac{1}{2}}(\ket{\mathsf  3}+\ket{\mathsf  4})$ and rewrite our Hamiltonian in the $\{\ket{\mathsf  1}, \ket{\mathsf  2}, \ket{+} \}$ space as:
\be
H_{Y4}^\prime = J_1\ket{\mathsf 1}\bra{\mathsf  2}+J_2\sqrt{2}\ket{\mathsf  2}\bra{\mathsf +} + H.c.
\ee
This shows that our system is equivalent to a 1D three-site spin chain where the ``output''
site is the symmetric entangled state
$(|0\rangle_3|1\rangle_4 +
|1\rangle_3|0\rangle_4)$. Such a chain effects perfect state transfer from the input site 1 to the
entangled output so long as the couplings are equal -- i.e. if $J_1 = J_2\sqrt{2}$.

\begin{figure}\bc
\includegraphics[scale=0.3]{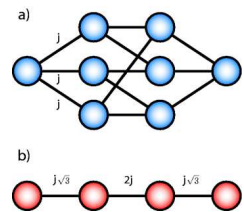}
\vspace{0.5cm}
\caption{(a) Example of a qubit network. All the couplings are equal and have strength $j$. (b) Equivalent one-dimensional chain with unequal couplings.}
\label{projection}
\ec\end{figure}

In the following we will show that we can extend this entanglement creation and distribution to
spin chain systems with
longer arms. To do this using the approach introduced above is rather cumbersome. Rather, we shall use the method of analysis introduced by Christandl {\it et al.} in \cite{chr05}. They find that perfect transfer along a longer chain is possible through the use of $N-1$ {\it unequal} couplings along an $N$ site chain, which satisfy
\be
J_{i,i+1} = \alpha \sqrt{i(N-i)} .
\label{coupling}
\ee
where $\alpha$ is a constant that sets the size of the interaction and $J_{i,i+1}$ is the coupling between site $i$ and site $i+1$. The authors of  \cite{chr05} also show that an array of spins with  equal couplings $j$ can be projected onto a 1D chain of unequal couplings. In particular, consider an array that is arranged as a series of columns. Let each node of column $i$ be connected to $n$ different nodes of adjacent column $i+1$, and each node of column $i+1$ be connected to $m$ different nodes of column $i$. Christandl {\it et al.} then show that each column can be projected onto a single node of a one dimensional spin chain with coupling $j\sqrt{mn}$ between nodes $i$ and $i+1$. An example is shown in figure~\ref{projection}.

If we analyze the four site Y-chain of figure 1 in this way, it is trivial to see that the Y-chain is equivalent to a 1D chain with couplings $J_1, J_2\sqrt{2}$, exactly as before.
According to the general properties of a 3-node chain~\cite{chr05}, this structure allows for perfect transfer between the two extremes so long as $J_1 = J_2\sqrt{2}$. In this case, the perfect transfer of an excitation in node 1 is allowed across the Y-shaped structure to the sites at the the extremes 3 and 4, where it must be shared between the two ends, giving rise to entanglement. 
Interestingly, the same $\sqrt{2}$ hub branching factor prevents hub reflection for propagating wave packets in divided chains of harmonic oscillators~\cite{perales05}, as well as Gaussian wave packets of propagating electrons or magnons~\cite{yang06, yang07}.

\begin{figure}\bc
\includegraphics[scale=0.35]{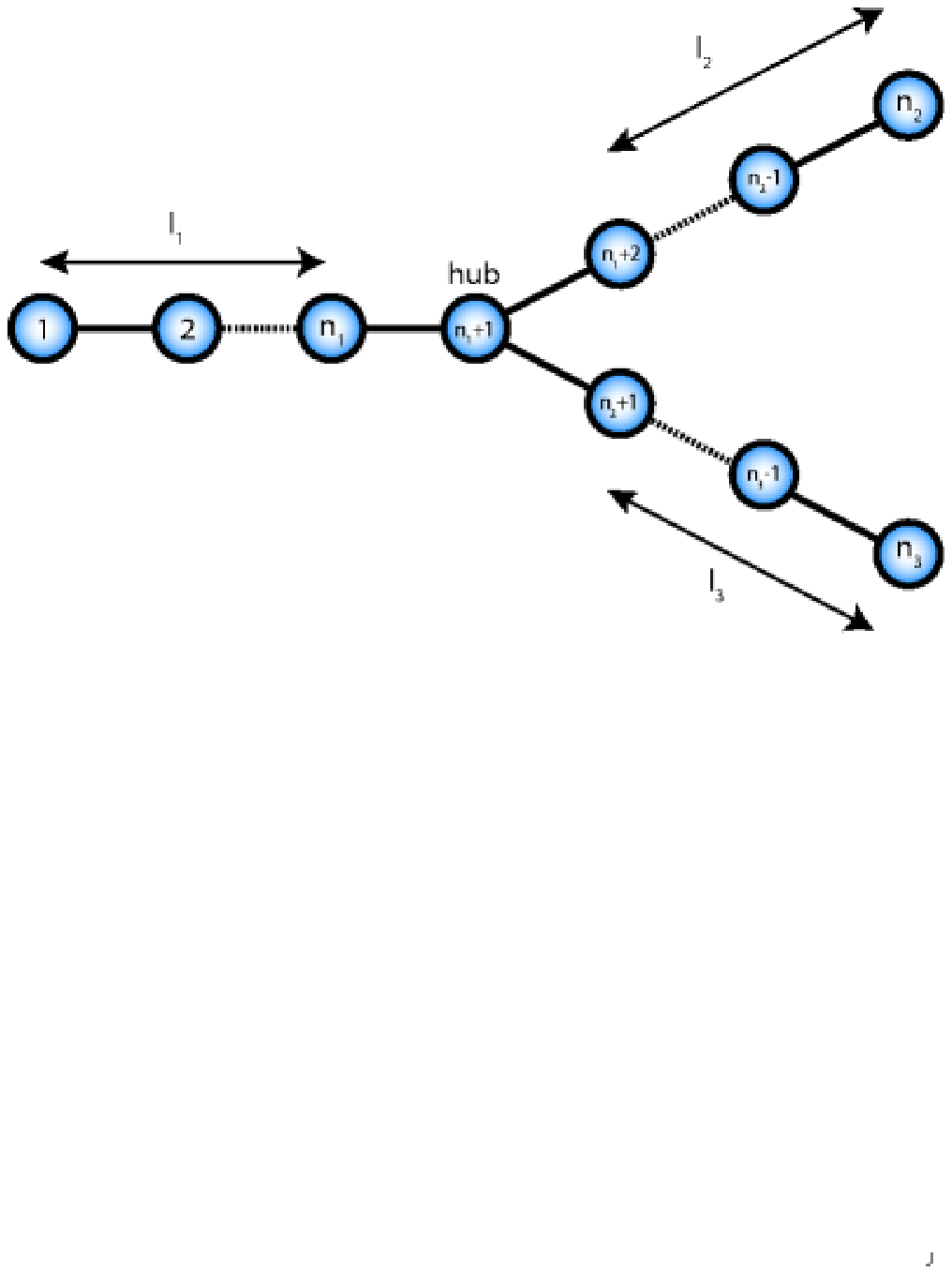}
\vspace{-5cm}
\caption{Relevant notation for `star' structures.}
\label{fig3}
\ec\end{figure}

We can extend this kind of analysis to larger structures, and we now introduce a labelling convention to deal with these. Referring to figure~\ref{fig3}, we choose to number branches clockwise,
starting from the input chain. The length of branch $k$ (excluding the hub)
is indicated as $l_k$. A Y-structure is then specified by the
sequence of numbers $(l_1,l_2,l_3)$.  Integer labels $i$ will be used for each of the total $N$ sites in the structure as follows: Starting with site $i=1$ at the left-most site of the input chain, label in sequence up to and including the hub. Continue by labelling along output branch 2, to its end, and then continue further by labelling branch 3, starting from the site adjacent to the hub and going to its end.
If the site furthest from the hub of branch $k$ is called $n_{k}$ then
$n_{1}=1$, $n_2=l_1+l_2+1$ and $n_{3}=N=l_1+l_2+l_3+1$.

\begin{figure}\bc
\includegraphics[scale=0.32]{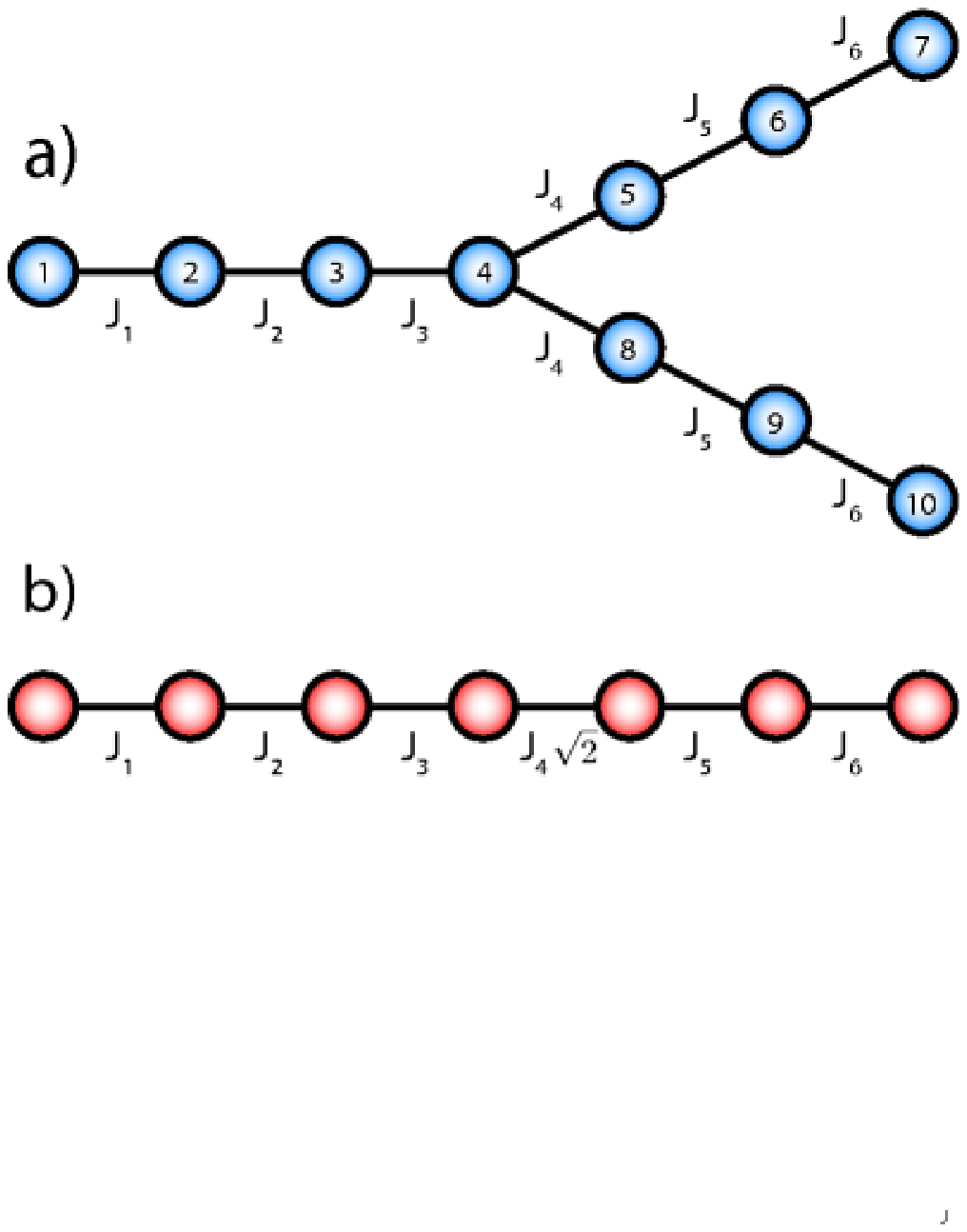}
\vspace{-3cm}
\caption{(a) Ten site Y spin chain network. Its equivalent one-dimensional representation is shown in (b).}
\label{10site}
\ec\end{figure}

\begin{figure}\bc
\includegraphics[scale=0.42]{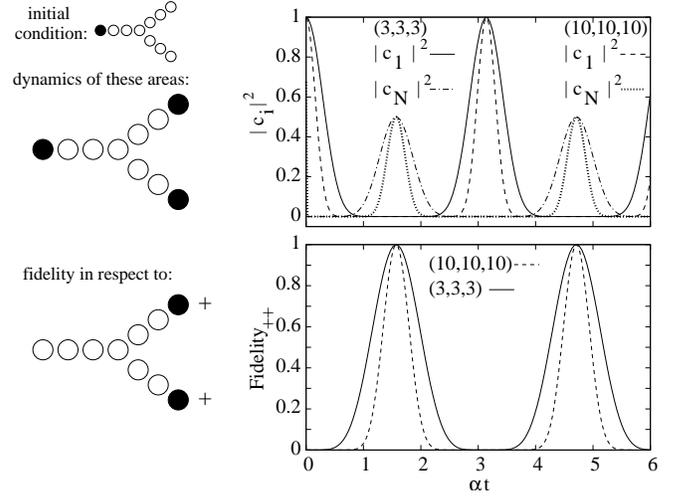}
\caption{Results of simulations on a (3,3,3) and a (10,10,10) Y spin chain system,
with nearest neighbour couplings chosen to satisfy
 (\ref{coupling}) and the hub branching rule. Initial condition: $c_1=1$
Upper panel: $|c_1|^2$ and $|c_N|^2$ with respect to the rescaled time $\alpha t$.
Lower panel: corresponding fidelity with respect to the state
$\ket{+}\equiv 2^{-\frac{1}{2}}(\ket{\mathsf  n_{2}}+\ket{\mathsf  n_{3}})$.
}
\label{fig4}
\ec\end{figure}

To demonstrate the
generation and perfect transfer of bi-partite
entanglement to the output spins $n_{2}$ and $n_{3}$, we have  performed
dynamical simulations using the Hamiltonian of  (\ref{Hspin}),
the condition  (\ref{coupling}) and the branching rule
for the couplings at the hub spin. Figure~\ref{10site} shows  how the hub branching rules applies to the (3, 3, 3) structure, where
$J_1=J_6=\alpha\sqrt{6}, J_2 = J_5 = \alpha\sqrt{10},
J_3 = \alpha\sqrt{12}, J_4 = \alpha\sqrt{6}$ for perfect state transfer.
Figure~\ref{fig4} shows the result of the simulations for both the (3, 3, 3), $N=10$, and (10, 10, 10), $N=31$, structures.
The initial condition is $c_1=1$ (and all others zero),
where $c_i$ indicates the amplitude coefficient
for the state $\ket{i}$.
The upper panel of figure~\ref{fig4} shows the temporal evolution of $|c_1|^2$ and
$|c_N|^2$ for the two systems. We underline that, due to symmetry,
spins $N(=n_3)$ and $n_2$ have the same dynamics in this case, and our results show that the excitation is completely transferred from site $1$ to sites $n_2$
and $n_3$ at time $\pi/2 \alpha$ and periodically returns there at regular time intervals of $\pi/\alpha$.
In respect to the rescaled time $\alpha t$, the peak corresponding
to each revival narrows as the length $l_k$ of the branches
increases. This is important when considering an
optimal practical structure for distributing entanglement, since
a narrower peak puts greater time constraints on any entanglement extraction protocol.
The lower panel of figure~\ref{fig4} shows the corresponding fidelity with respect to the
state
$\ket{+}\equiv 2^{-\frac{1}{2}}(\ket{\mathsf  n_{2}}+\ket{\mathsf  n_{3}})$.

We can quantify the amount of entanglement at the chain ends (as a
function of time) more formally by using the entanglement of
formation, $E_F$. This has a value unity  for a maximally entangled
state and zero for an unentangled one. Specifically, it measures the
number of Bell states required to create the state of interest and
for a two qubit state it is given by:
\begin{equation}
E_F({\rho})=h\left(\frac{1+\sqrt{1-\tau}}{2}\right),
\end{equation}
where $h(x) = -x\log_2(x) - (1 - x)\log_2(1 - x)$ is the Shannon
entropy function. $\tau$ is the ``tangle'' or ``concurrence'' squared:
\begin{equation}
\tau={\cal
C}^{2}=\left[\max\{\lambda_1-\lambda_2-\lambda_3-\lambda_4,0\}\right]^{2}.
\end{equation}
The $\lambda$'s are the square roots of the eigenvalues, in
decreasing order, of the matrix
${\rho} \tilde{{\rho}} = {\rho}\;\sigma_{y}^{A}
\otimes \sigma_{y}^{B} {\rho}^{*} \sigma_y^A \otimes
\sigma_{y}^{B}$, where ${\rho}^{*}$ denotes the complex conjugation
of ${\rho}$
in the computational basis $\ket{00}, \ket{01}, \ket{10}, \ket{11}$~\cite{munro01}.

\begin{figure}\bc
\includegraphics[scale=0.5]{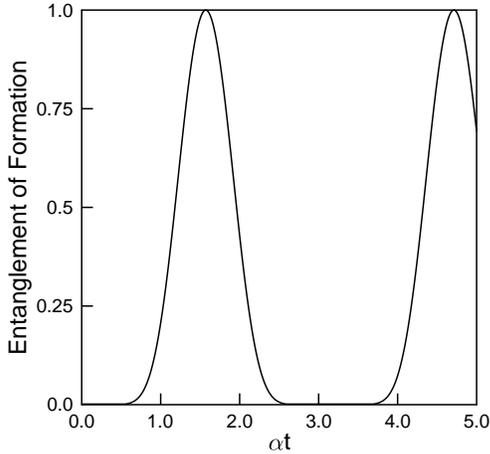}
\vspace{-1cm}
\caption{Entanglement of Formation of the two chain end qubits ($n_2 = 7$ and $n_3=10$) of the ten site Y chain shown in figure~\ref{10site}, following the initialization procedure described in the text.}
\label{eof}
\ec\end{figure}

$E_F$ between qubits $n_2=7$ and $n_3=10$ of the (3, 3, 3) structure is shown in figure~\ref{eof}, as a function of time following initialization of the excitation on site 1. As would be expected, a maximally entangled
state is obtained after a time $\pi/2\alpha$ and at intervals of $\pi/\alpha$ thereafter.

\begin{figure}\bc
\includegraphics[scale=0.43]{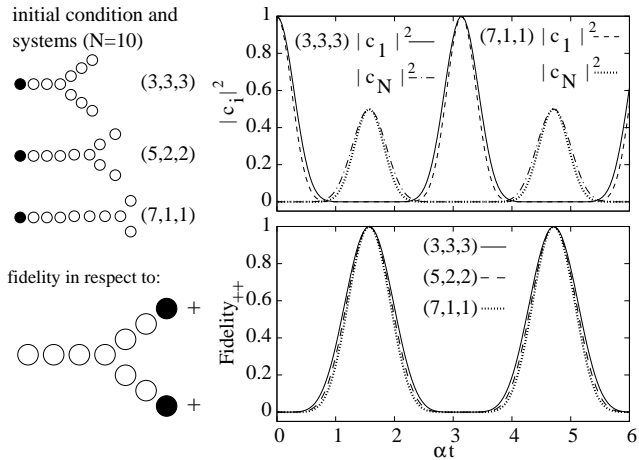}
\caption{Results of simulations on a $(3,3,3)$, $(5,2,2)$ and
$(7,1,1)$ Y spin chain system, with nearest neighbour couplings
chosen to satisfy  (\ref{coupling}) and the hub branching rule.
Initial condition: $c_1=1$ Upper panel: $|c_1|^2$ and $|c_N|^2$ as a function of
the rescaled time $\alpha t$. Lower panel: corresponding
fidelity with respect to the state $\ket{+}\equiv
2^{-\frac{1}{2}}(\ket{\mathsf n_2}+\ket{\mathsf  n_3})$ }
\label{fig42}
\ec\end{figure}

Maximally entangled states can be produced even if the length of
the output branches, $l_2=l_3$, is different from the length of
the input branch $l_1$, as long as the nearest neighbour couplings  satisfy
 (\ref{coupling}) and the  hub branching rule.
As an example in figure~\ref{fig42} we compare the results for the systems
(3, 3, 3), (5, 2, 2) and (7, 1, 1), with notation and initial condition as in figure~\ref{fig4}.
Our results show the perfect generation
of an output entangled state, as before, at time 
$t=\pi/2\alpha$ from a simple initial input excitation
even for the more asymmetric systems. As the asymmetry increases,
the width of the peaks decreases, but the change is not significant.

\section{Generalization to $p$ outputs}
\begin{figure}\bc
\includegraphics[scale=0.3]{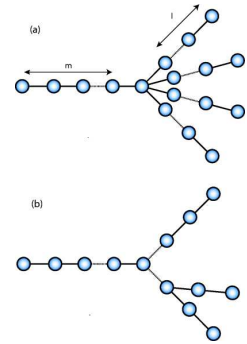}
\caption{(a) Example of a star-like family member $(m,l,l,l,l)$.
(b) Example of a bifurcation-like family member.
}
\label{n_out}
\ec\end{figure}

The branching rule we have given for preparation of bi-partite
entanglement using a Y spin chain structure can be extended to the
case of the different families of tree-like spin chain structures
with a single input and $n$ output branches. Examples of their
members are  represented in  figure~\ref{n_out}.

Panel (a) corresponds to a `star-like' family, in which there is just one hub
and all output branches have the same length. Extending the notation early introduced,
we can indicate the members of this family  as $(m,l,l,l,...,l)$.
Following on from the arguments in Section~\ref{bipartite}, the coupling
on the outputs from the hub spin are
all reduced by a factor of $1/\sqrt{p}$ compared to the effective
one-dimensional spin chain values. (Again, an analogous factor also exists for the condition of no hub reflection for a wave packets excitation on the bosonic or propagating fermion systems discussed previously~\cite{perales05, yang06, yang07}.)
The entangled state which can be created by using this family is a W-state,
consisting of an excitation shared between $p$ separated sites, with the
 {\it symmetric} form $(\ket{1_{n_2},0_{n_3},...,0_{n_p}}+\ket{0_{n_2},1_{n_3},...,0_{n_p}}+...+\ket{0_{n_2},0_{n_3},...,1_{n_p}})/\sqrt{p}$.
In real systems it is likely to be
impractical to prepare $p$-way entangled states using this approach
for large values of $p$. However, if three-dimensional physical structures
can be built--- and for example there is the potential for this with quantum dot
systems---modest values of $p>2$, such as 3, 4 and 5 could be possible.

Panel (b) gives an example of what we refer to as a bifurcation-like structure.
These structures contain more than one hub -- and for perfect transfer the branching rule must be implemented at each. Note that for timing of the dynamical evolution
to produce complete excitation transfer to the outlying sites,
the number of sites along all output paths from the initial hub must be the same.
The entangled states which can be
created by using this family share the excitation between the
different separated outlying sites with different unequal weights. e.g. 
the structure represented in panel (b) gives the
form $\ket{1_{n_2},0_{n_3},0_{n_4}}/\sqrt{2}+\ket{0_{n_2},1_{n_3},0_{n_4}}/2+\ket{0_{n_2},0_{n_3},1_{n_4}}/2$.
With only modest branching at each hub, a bifurcating structure
potentially could be fabricated in a planar arrangement. This could
in principle give rise to a significant number of output branches, whose
outlying spins can be entangled, sharing an excitation in an {\it asymmetric} manner,
where the type of asymmetry depends on the number of hubs and on their position.

\section{Extraction of entanglement}
Having generated and distributed, for example, bi-partite entanglement
using a Y spin chain system, it is useful to be able to isolate this
entanglement in some way, so it can be used as a resource. Clearly,
as illustrated in figure~\ref{eof},
the arrival time of the entangled state at the output
relative to the preparation time of the excitation at the input is
known. However, if the entanglement is being created as a resource (and particularly
if a number of such states are to be purified before use)
then it is essential to preserve it once it is created.
This might be done by extracting it (e.g. with swap operations)
from the spin chain system and
transferring it to a storage system of qubits, or to
members of spatially separated quantum registers or processors.
See  \cite{spi06}
for a method of doing this with an excitonic quantum dot system.
Certainly such swapping is likely to be necessary in any excitonic realization of a
spin chain, if the entangled resource is required to exist on timescales
longer than the ground state exciton coherence times and lifetimes\cite{spi06}.
Another possibility is to physically isolate the two nodes at the end of the chain,
after the entanglement has formed there.   If fast local
control of the couplings is possible, then the state could indeed be
frozen at the ends of the output branches.
To achieve this the couplings have to be switched fast on a timescale set by
the chain dynamics, as illustrated by figure~\ref{eof}, and this has to be
done simultaneously on both output branches.
In general both the freezing of the spatially separated entanglement, and its
swapping into longer-lived ``memory'' qubits at the ends of the output
branches of the spin chain systems, requires fast, coordinated action at
the ends of both branches. This may be fine if these actions are both
controlled through external optical pulses.

However, there is another approach to preserving the entanglement,
which could be very effective for some realizations of the spin chain systems.
This can be implemented when an entangled state is
(a) an eigenstate of the Hamiltonian  (\ref{Hspin}) and
(b) it can be reached with just {\it one} single qubit local operation
from the entangled state provided by the natural dynamics of the system.
In this case isolation of the entanglement requires just a single fast
single-qubit operation at one of the two branch ends.
The simplest example
of this is again with the four-site system illustrated in figure~\ref{fig1}. As
already remarked,  the antisymmetric entangled state
$\ket{-}\equiv2^{-\frac{1}{2}}(\ket{\mathsf  3}-\ket{\mathsf  4)}$
is an eigenstate of the system, decoupled
from the transfer dynamics (The state of qubits 3 and 4 for a
system in the $\ket{-}$ state, after tracing out the others, is
the maximally entangled
$2^{-\frac{1}{2}}(|0\rangle_3|1\rangle_4 - |1\rangle_3|0\rangle_4)$).
After state transfer, our system is in the $\ket{+}$ state defined
earlier. If a phase flip gate, for example
$|0\rangle \rightarrow |0\rangle$ and
$|1\rangle \rightarrow -|1\rangle$,
fast on the timescale of the spin chain dynamics, is applied to
just one chain end qubit (3 or 4) of the $\ket{+}$ state,
we obtain $\ket{-}$. Thus our entangled resource can be frozen
into an eigenstate of this four-site system.

\begin{figure}\bc
\includegraphics[scale=0.45]{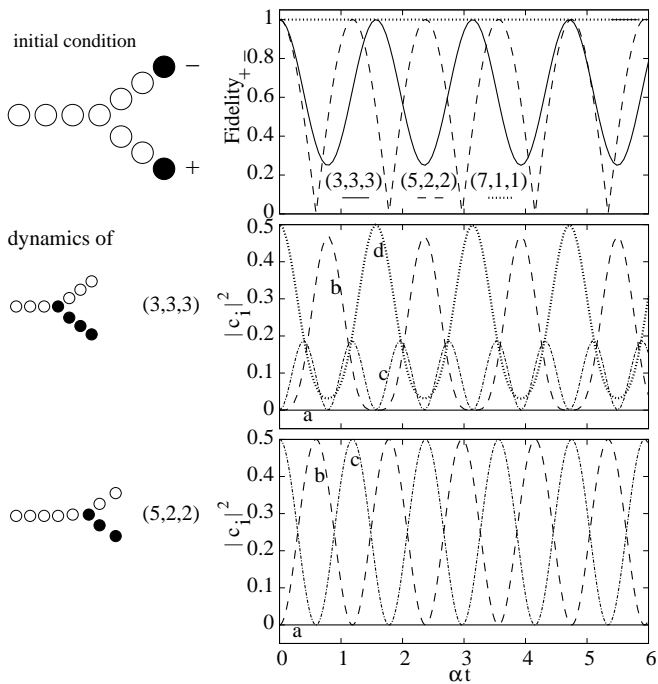}
\caption{Simulations on a $(3,3,3)$, $(5,2,2)$ and $(7,1,1)$ Y spin
chain system, with nearest neighbour couplings chosen to satisfy
 (\ref{coupling}) and the hub branching rule. Initial condition:
$\ket{-}\equiv 2^{-\frac{1}{2}}(\ket{\mathsf  n_2}-\ket{\mathsf
n_3})$. Upper panel: Fidelity with respect to the  state $\ket{-}$
as a function of the rescaled time $\alpha t$. Middle panel:  $|c_i|^2$
for the spins in $l_3$ plus the hub spin for $(3,3,3)$. The line
labelled as `a' represents the hub dynamics; `b' to `d' represent the
dynamics of spins in $l_3$ from the closest to the hub (b) to
$n_3$ (d). Lower panel: $|c_i|^2$ for spins in $l_3$ plus the hub
for $(5,2,2)$. The line labelled as `a' represents the hub dynamics,
`b' the dynamics of the spin closest to the hub and `c' the dynamics
of $n_3$. } \label{fig5}
\ec\end{figure}

\begin{figure}\bc
\includegraphics[scale=0.45]{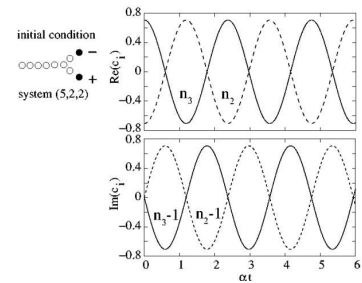}
\caption{Results of simulations on a $(5,2,2)$ Y spin chain system,
with nearest neighbour couplings chosen to satisfy
 (\ref{coupling}) and the hub branching rule. Initial condition:
$c_{n_3}=-c_{n_2}=1/\sqrt 2$
Dynamics of real part of $c_{n_2}$ and $c_{n_3}$ (Upper panel)
and imaginary part of $c_{n_2-1}$ and $c_{n_3-1}$ (Lower panel).
}
\label{fig52}
\ec\end{figure}

Things are not quite so simple in larger Y spin chain systems. If
a phase flip is applied to the output entangled state of a larger Y
spin chain system, this opposite parity entangled state is not an
eigenstate of the full system. However, the entanglement is
effectively ``dynamically frozen'' by the action of the single-qubit
phase flip, in the sense that the subsequent dynamics just involves
the spins in $l_2$ and $l_3$ and the $\ket{-}$ entangled state
between spins $n_2$, $n_3$ revives periodically and more
frequently than if the phase flip is not applied.  This is
illustrated in figure~\ref{fig5} for the systems $(3,3,3)$, $(5,2,2)$
and $(7,1,1)$, where the results of simulations are presented for
the initial condition $\ket{-}\equiv 2^{-\frac{1}{2}}(\ket{\mathsf n_2}-\ket{\mathsf  n_3})$. 
Couplings are chosen to satisfy  (\ref{coupling}) and the hub branching rule.
The upper panel shows the fidelity
with respect to the state $\ket{-}$ (`Fidelity$_{+-}$'). These
results show the perfect revival
of an output entangled state with a period shorter than
$\pi/\alpha$. For the system $(3,3,3)$, comparison between
figure~\ref{fig42} and figure~\ref{fig5} shows that the revival spacing
is $\pi/2\alpha$ in this case. As the dynamics of the system is
restricted by quantum destructive interference to the spins in $l_2$
and $l_3$, the time of revival decreases with decreasing $l_2=l_3$.
For the limit case $(7,1,1)$ -- in which $l_2=l_3=1$ -- the state
$\ket{-}$ is again an eigenstate of the system, so that the
entangled state is frozen at the output spins $n_2$ and
$n_3$. This suggests that a system represented by $(m,1,1)$
 could be used to create
 entanglement in two qubits on a distant ($m$ spins away) array or register. In this case
 the entanglement could then be trapped in the two qubits by a
 single qubit operation applied to one of them.
This system could also be used as a switch, for example to start a
computation, or to initialize from a distance the creation of
entangled states needed for measurement-based computation
\cite{dev05}. The middle panel shows the dynamics of $|c_i|^2$ for
the spins in $l_3$ plus the hub spin for $(3,3,3)$. Clearly the hub
is never involved in the dynamics  (the line labelled as `a'). The
lines labelled from `b' to `d' correspond to the dynamics of spins in
$l_3$ from the hub outward. No spin in $l_1$ is involved in the
dynamics (not shown). This can be understood from initial condition,
in which the dynamics of the two output branches is out of phase by
$\pi$. The symmetry of the structure and of the bonds must therefore
generate {\it destructive} interference in $l_1$ and the hub. The
lower panel shows the corresponding dynamics for the $(5,2,2)$ system.
Note that, since in this case $l_2=l_3=2$, the dynamics oscillates
perfectly between the spins $\{n_2, n_3\}$ and their {\it
respective} nearest neighbours  $\{n_2-1, n_3-1\}$.
Figure~\ref{fig52} shows the detailed dynamics of the components of
$c_i$ for $\{n_2 , n_3\}$ (upper panel, where the coefficient is purely real) and
$\{n_2-1, n_3-1\}$ (lower panel, where the coefficient is purely imaginary). A
(maximally) entangled state of the type $\ket{-}$ is periodically and
perfectly transferred between the two pairs of spins. 
For the general $(x,2,2)$ system with this initial condition, the dynamics
oscillate between two maximally entangled states, which have a phase difference.

\begin{figure}\bc
\includegraphics[scale=0.42]{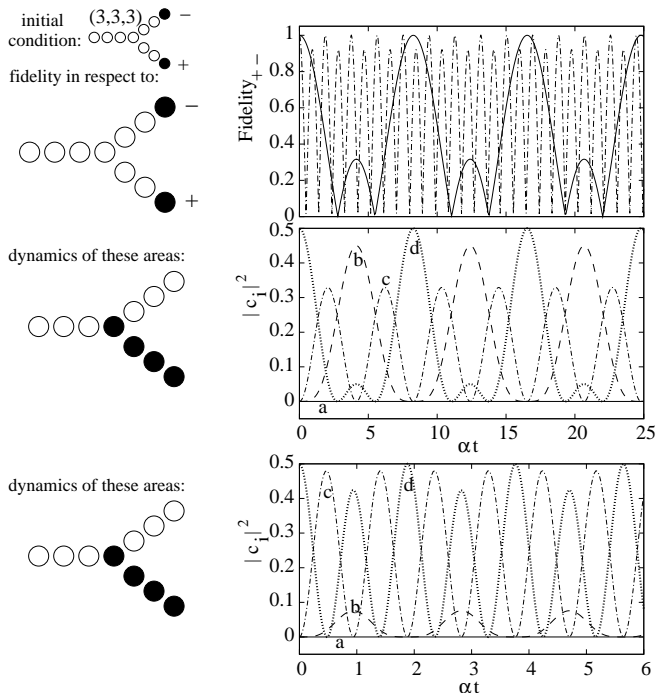}
\caption{Results of simulations of $(3,3,3)$ Y spin chain systems
with nearest neighbour couplings generated using a flat distribution over (0,1],
but satisfying the condition that the bonds in $l_2$ and $l_3$ (and to the hub) have the same sequence.
 Initial condition $\ket{-}\equiv 2^{-\frac{1}{2}}(\ket{\mathsf  n_2}-\ket{\mathsf  n_3})$.
Upper panel: Fidelity with respect to the state $\ket{-}$.
Middle panel: $|c_i|^2$ versus  the rescaled time $\alpha t$ for
spins in $l_3$ plus the hub for the random set of couplings A (solid line in upper panel).
The line labelled as `a' represents the hub dynamics, and `b' to `d'
the dynamics of spins in $l_3$ from the closest to the hub (b) to $n_3$ (d).
Lower panel: As for the middle panel but  for random set B (dashed-dot line in upper panel). Note the different scale on the time axis here.
}
\label{fig53}
\ec\end{figure}

As pointed out above, the destructive interference at the hub and
in branch $l_1$ is caused by (i) the antisymmetric initial condition of the type
$\ket{-}$, (ii) the symmetry of the systems, i.e. $l_2=l_3$ and
(iii) the same sequence of bonds in  $l_2$ and $l_3$. What is more interesting
is that this is a {\it sufficient} condition to ensure perfect
periodic revival of entanglement at the output spins. This can be
understood by defining an operator $S$, which swaps the two outgoing
branches of any Y structure. Clearly $\ket{+}$ ($\ket{-}$) is an
eigenstate of $S$ with eigenvalue $+1$ ($-1$) and, further, $S$ commutes
with any Hamiltonian that has equivalent outgoing branches. We underline that these
need not satisfy any special coupling rule, only the condition that the two branches
have to be the same. $S$ is thus conserved and so any state with
eigenvalue $-1$ must always have zero amplitude at the hub and in
the input branch. For a finite Y system, an initial state of
$\ket{-}$ will therefore always revive, with the hub and input
branch having zero amplitude at all times. This can be illustrated
in numerical examples. In figure~\ref{fig53} we present the results of
simulations on a $(3,3,3)$ spin chain, for two systems in which the
set of nearest neighbour couplings is generated using a flat 
distribution in the interval (0,1], but
satisfying the condition that bonds in $l_2$ and $l_3$
(including the one to the hub) have the same sequence. The initial
condition is given by $\ket{-}\equiv 2^{-\frac{1}{2}}(\ket{\mathsf
n_2}-\ket{\mathsf n_3})$. The upper panel shows the fidelity
with respect to the state $\ket{-}$ for two choices of random couplings: A (solid line) and B
(dashed-dot line) as a function of the rescaled time $\alpha t$. As
can be seen, the fidelity pattern is more complicated than for the
case of coupling following the prescription given by
 (\ref{coupling}) and the hub branching rule. In particular peaks
are present which do not correspond to maximum fidelity.
Nevertheless peaks corresponding to fidelity of unity appear
periodically, with the period determined by the particular
sequence of bonds. The middle (lower) panel shows the dynamics of
$|c_i|^2$ for spins in $l_3$ and the hub for system A (B). 
We remark that neither system A nor B allows for perfect transfer between spin 1 and spins $\{n_2,n_3\}$: the condition for perfect
transfer is a more delicate one and in particular does not rely on
simple symmetry considerations.

\subsection{{\bf \it Complete} Freezing of Dynamically 
Generated Entanglement By Local Operations}

\begin{figure}\bc
\includegraphics[scale=0.45]{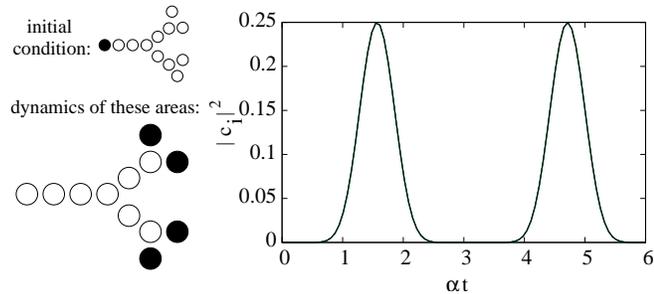}
\caption{Dynamics of $|c_{n_i}|^2$  (for the output branch end spins) versus
the rescaled time $\alpha t$, for the bifurcation structure shown. This follows the initialization condition $c_1 = 1$.
}
\label{figbifent}
\ec\end{figure}

Rather than simply dynamically freezing entanglement by shortening
its revival time, a combination of the entanglement extraction ideas
can be used to actually freeze spatially separated entanglement in
systems of the bifurcation family. For example, consider a Y
structure where at the end of each output branch there is a
bifurcation into two final spins. This is illustrated in
figure~\ref{figbifent}. In terms of the four end spins, the
natural dynamics will generate a state of the form
$\frac{1}{2}(\ket{0,0,0,1}+\ket{0,0,1,0}+\ket{0,1,0,0}+\ket{1,0,0,0})$.
This is shown in figure~\ref{figbifent} by plotting 
the branch end spin probabilities  $|c_{n_i}|^2$ as a function of rescaled
time $\alpha t$. If at a probability peak a phase flip is applied to
one spin out of each pair, a state of the form
$\frac{1}{2}(\ket{0,0,0,1}-\ket{0,0,1,0}+\ket{0,1,0,0}-\ket{1,0,0,0})$
results. This is an eigenstate of the system and the entanglement is
thus frozen. Although four spins are involved, the spatial
separation achieved is between two pairs of spins, so the pairs
could each be viewed as (singlet) storage buffers for one qubit. In this sense
the system contains {\it spatially separated and stored bipartite entanglement},
which could be released for future use by single qubit operations
and/or coupling to other systems at the branch ends.

\section{Conclusions}
We have shown how branching spin chain systems can be used
to both generate and distribute
entanglement from their natural dynamics.  Once distributed,
this entanglement can be isolated
through mapping the entanglement into specific qubits at the ends of branches, or,
as we have shown for distributed bipartite entanglement,
applying a simple single-qubit
operation to one end spin of a branch. Given its simplicity, this
latter approach may be particularly
useful in initial experimental investigations of such branched systems.

Distributed  entanglement provides a useful resource,
for example for teleportation~\cite{tele93} or distributed quantum processing.
In contrast to the use of spin chains to propagate quantum states
from one place to another with as high a fidelity as possible, there
could be some advantage in building up a high fidelity entangled resource
``off line''. Real systems, with their inevitable imperfections, will
almost certainly degrade transmission fidelities, even if in principle
these approach unity.
Certainly, with the ``off line'' resource approach,
purification~\cite{pur96} could be applied to
build up a higher fidelity resource than can be achieved by direct
transmission. This can then be used to transfer quantum states or
some form of quantum communication. In effect, the concept of a
quantum repeater \cite{repeater98} could be employed in a solid state,
spin chain scenario.

It is clear from the oscillatory behaviour  of the dynamics that
branched spin chain systems could be operated ``in reverse'', for
example providing a method of detecting certain entangled states. If
an entangled state is fed into the ends of branches it can deliver a
signature at a certain single node or spin (such as the one we have
used as the input node in our entanglement creation scenarios). This
could be detected with just a single-qubit measurement. In more
involved scenarios, entangled states could also be prepared
non-locally, through a simple conditioning measurement on one or
more spins, in conjunction with the natural spin chain dynamics
acting on some simple initial state.

Finally, we comment that all these related effects result from the
basic dynamics of the branched spin chain systems and simply
prepared initial states. Whilst some control is need over the
couplings to achieve entanglement creation, distribution and
isolation, there is certainly significant potential for branched
spin chain systems in solid state quantum processing and
communication. This potential will continue to grow, as fabrication
or creation of solid state systems that can operate as spin chains
continues to progress.

BWL acknowledges support from the Royal Society through a University Research Fellowship, the QIPIRC (www.qipirc.org, GR/S82176/01),
DSTL and St Anne's College, Oxford. IDA acknowledges support from the
Dept. of Physics, University of York,  and the kind hospitality of Hewlett Packard Labs, Bristol,
and the Department of Materials, University of Oxford.


\end{document}